\documentclass[submit]{eps}

\usepackage{graphicx}
\usepackage{times}

\volumenumber{63}
\publishyear{2011}
\frompage{\pageref{firstpage}}
\topage{\pageref{finalpage}}

\shorttitle{Nisha Katyal \lowercase{\textit{et al.}}: Interstellar Grains}

\title{Interstellar Grains: Effect of Inclusions on Extinction}
\author{Nisha Katyal$^1$, Ranjan Gupta$^1$ and D. B. Vaidya$^2$}
\affiliation{$^1$IUCAA, Post Bag 4, Ganeshkhind, Pune-411007, India\\
             $^2$ICCSIR, Ahmedabad 380006, India}
\received{September 8, 2010}\revised{June 16, 2011}\accepted{June 17, 2011}
\abstract{

A composite dust grain model which simultaneously explains
the observed interstellar extinction, polarization, IR emission
and the abundance constraints, is required.
We present a composite grain model, which is made up
of a host silicate oblate spheroid and graphite inclusions.
The interstellar extinction curve is evaluated in the spectral
region 3.4-0.1$\mu m$ using the extinction efficiencies of the
composite spheroidal grains for three axial ratios.
Extinction curves are computed using the discrete dipole approximation (DDA).
The model curves are subsequently compared with the average
observed interstellar extinction curve and with an extinction curve
derived from the IUE catalogue data.}


\begin{document}
\label{firstpage}
\maketitle
\copyrighttext{}

\section{Introduction}

It is highly unlikely that the interstellar grains are spherical in shape
or that they are homogeneous in composition and
structure. The collected interplanetary particles are nonspherical and highly
porous and composites of very small sub-grains glued together (Brownlee,
1987). The existence of interstellar polarization requires that the
interstellar grains must be nonspherical.
The elemental abundances derived from the observed
interstellar extinction also do not favour the homogeneous
composition for the interstellar grains. There is no exact theory
to study light scattering by inhomogeneous grains (viz. porous, fluffy
and composite).
We have used Discrete Dipole Approximation (DDA) to study
the extinction properties of the composite grains. For the
description on the DDA see Draine (1988).
In the present study, we calculate the extinction efficiencies
for the composite oblate spheroidal grains, made up of the host silicate
spheroid with
embedded inclusions of graphite, in the wavelength region,
3.4-0.10 $\mu m$.  Using these extinction efficiencies of the composite grains with a
power law grain size distribution we evaluate the interstellar extinction
curve. We also estimate the cosmic abundances viz. silicon and carbon for the grain
models which fit the observed interstellar extinction curve.
It must be mentioned here that the composite oblate grain model presented in this study
has also been used to interpret the observed IR emission from circumstellar
dust (Vaidya \& Gupta, 2011).

In section 2 we give the validity criteria for the DDA and the composite oblate grain models.
In section 3 we present the results of our computations and discuss
them.
The main conclusions of our study are given in section 4.

\subsection{Composite grains and DDA}

In the Discrete Dipole Approximation (DDA), a solid particle is replaced (approximated)
by an array of N dipoles. When a grain is exposed to an electromagnetic wave, each dipole
responds to the radiation field of the incident wave as well as to the fields of the
other N-1 dipoles that comprise the grain (Draine, 1988).

We use the computer code developed by Dobbie (see Vaidya et. al., 2001) to generate the
composite oblate grain models used in the present study.
The constituent materials of the composite grains consist of silicates and graphites, since in
the interstellar medium, carbon and silicate occur separately and in the form of
small particles which are agglomerated into large grains. 
For detailed discussion on the compostion of the
composite interstellar dust see Mathis (1996) and Vaidya et. al. (2001).
We have studied composite grain models with a host silicate
spheroid containing N= 9640, 25896 and 14440 dipoles, each carved out from
$32 \times 24 \times 24$, $48 \times 32 \times 32$
and  $48 \times 24 \times 24$ dipole sites, respectively;
sites outside the spheroid are set to be vacuum and sites inside are
assigned to be the host material.
It is to be noted that the composite oblate spheroidal grain with
N=9640 has an axial ratio (AR) of 1.33, whereas N=25896 has the axial
ratio of 1.5 and N=14440 has the axial ratio of 2.0.
Further, if the semi-major axis and semi-minor axis are denoted by x/2 and y/2,
respectively, then $a^{3}=(x/2)(y/2)^{2}$, where a is the radius of the sphere whose volume is same as of a spheroid. To study randomly oriented spheroids, it is necessary to get the
scattering properties of the composite grains over all possible
orientations. We use three values of each of the orientation parameters ($ \beta, \theta
and \phi$). i.e. averaging over 27 orientations, which we find is quite adequate
(see Wolf et al., 1994).
The volume fractions of the graphite inclusions used are
10\%, 20\% and 30\% (denoted as f=0.1, 0.2 and 0.3).
The size of the inclusion is given by the number of dipoles, 'n' across the diameter of an 
inclusion; e.g. 152 for composite grain model with N=9640 i.e. AR=1.33 (see Table I in Vaidya
and Gupta, 2011).
Details on the computer code and the corresponding modification
to the DDSCAT 6.1 code (Draine \& Flatau 2003) are given in 
Vaidya et al. (2001) and Gupta et al. (2006).
The modified code outputs a three-dimensional matrix specifying
the material type at each dipole site; the sites are either silicate,
graphite or vacuum.
For an illustrative example of a composite oblate spheroidal grain
with N=14440 dipoles (AR=2.00), please refer Figure 1, given in
Gupta et al. (2006).
There are two validity criteria for DDA (see e.g. Wolff et al. 1994);
viz. (i) $\rm |m|kd \leq 1$, where m is the complex refractive index
of the material, k=$\rm \pi/\lambda$ is the wavenumber and
d is the lattice dispersion spacing and
(ii) d should be small enough (N should be sufficiently large) to
describe the shape of the particle satisfactorily.
The complex refractive indices
for silicates and graphite are obtained from Draine (1985, 1987).
For all the composite grain models, with N=9640, 25896 and
14440 (i.e. AR=1.33, 1.50 \& 2.00 respectively)
and for all the grain sizes, between a = 0.001-0.250$\mu$, in
the wavelength range of 3.4-0.1$\mu m$, considered in the present
study; we have checked that the DDA criteria are satisfied (Vaidya et al. 2007).

\section{Results}

\subsection{Extinction Efficiency of Composite Grains}

In the present paper, we study the extinction properties of the spheroidal 
grains with three axial ratios (AR), viz. 1.33, 1.5 and 2.0, corresponding to
the grain models with N=9640, 25896 and 14440 respectively,
for three volume fractions of inclusions; viz. 10\%, 20\% and 30\%,
in the wavelength region 3.4-0.10$\mu m$.
Figures 1 (a),(c) and (d) show the extinction efficiencies ($\rm Q_{ext}$) for the
composite grains with the host silicate spheroids containing N=9640,
25896 and 14440 dipoles, corresponding to axial ratio 1.33, 1.5
and 2.0 respectively with a host composite grain size set to a=0.01$\mu$.
The three volume fractions, viz. 10\%, 20\% and 30\%, of graphite
inclusions are also listed in the top (a) panel and an additional
volume fraction of 40\% is also displayed.
The extinction in the spectral region 0.28-0.20$\mu m$ is highlighted
in the panel (b) of this figure for the composite grains with N=9640.

\begin{figure}[t]
\centerline{\includegraphics[width=8.0cm,clip]{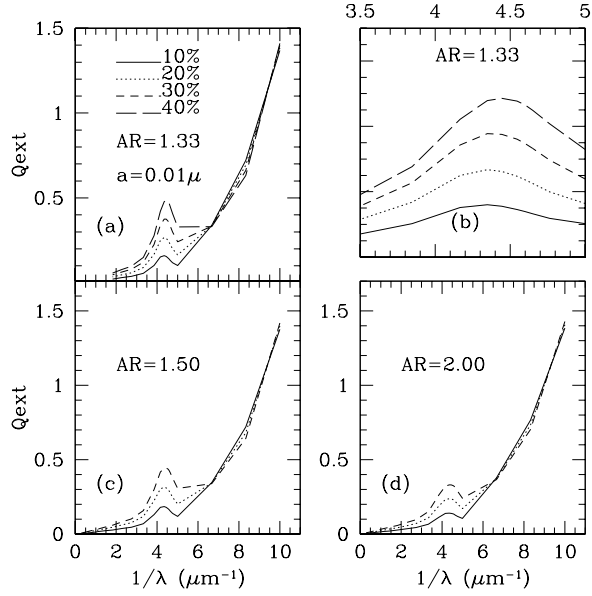}}
\caption{Extinction Efficiencies for the composite grains of size 0.01$\mu$
with host spheroids containing dipoles N=9640, 25896 and 14440; corresponding to
the axial ratio 1.33, 1.50 and 2.00 respectively, are
shown in (a),(c) and (d) in the wavelength region, 3.4-0.10$\mu m$ for three
graphite inclusion fractions (f=0.1, 0.2 and 0.3 corresponding to
10\%, 20\% and 30\% inclusion fractions). An additional 40\% inclusion case for
AR=1.33 (N=9640) is shown in the panels (a) and (b). In panel (b) the extinction curves in 
the wavelength region 0.28-0.20$\mu m$ are highlighted.}
\end{figure}

The effect of the variation of volume fraction of inclusions is
clearly seen for all the models. The extinction efficiency increases as
the volume fraction of the graphite inclusion increases.
It is to be noted that the wavelength of the peak extinction
shifts with the variation in the volume fraction of inclusions.
These extinction curves also show the variation in the width
of the extinction feature with the volume fraction of inclusions.
All these results indicate that the inhomogeneities within the
grains play an important role in modifying the '2175\AA~' feature.
Voshchinnikov (1990) and Gupta et al. (2005) had found
variation in the '2175\AA~' feature with the shape of the grain, and
Iati et al. (2001, 2004); Voshchinnikov (2002);
Voshchinnikov and Farafanov (1993) and Vaidya et al. (1997, 1999)
had found the variation in the feature with the porosity of the grains.
Draine \& Malhotra (1993) have found relatively little effect on either
the central wavelength or the width of the feature for the coagulated
graphite silicate grains.
Figures 2(a-d) show the extinction efficiencies ($\rm Q_{ext}$) for the
composite grains for four host grain sizes: viz. a=0.01, 0.05,
0.1 and 0.2 $\mu$ at a constant volume fraction of inclusion of 20\%.
It is seen that the extinction and the shape of the extinction curves
varies considerably as the grain size increases. The '2175\AA~ feature'
is clearly seen for small grains ; viz. a=0.01 and
0.05$\mu$, whereas for larger grains the feature almost disappears.

\begin{figure}[t]
\centerline{\includegraphics[width=8.0cm,clip]{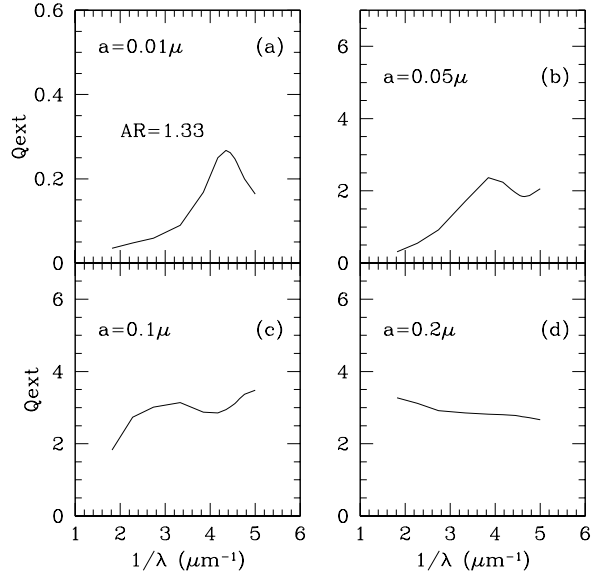}}
\caption{Extinction efficiencies for the composite grains with
AR=1.33 (N=9640) and with 20\% volume fraction of graphite inclusions for various grain
sizes.}
\end{figure}

\subsection{Interstellar Extinction Curve}

The interstellar extinction curve (i.e. the variation of extinction
with wavelength) is usually expressed by the ratio
$\rm E(\lambda - V)/E(B - V)$ versus 1/$\lambda$.
We use the extinction efficiencies of the composite grains,
with a power law size distribution (i.e. $\rm n(a) \sim a^{-3.5}$,
(Mathis et. al 1977) to evaluate the interstellar
extinction curve in the wavelength region of 3.4-0.10$\mu m$.
In addition to the composite grains, a separate component of small
graphite grains is required to produce the observed
peak at 2175\AA~ in the interstellar extinction curve (Mathis, 1996).
The stability of the bump at 2175\AA~ along all the lines of sight
rules out the possibility of using just composite grains, made up of silicate
with graphite as inclusions, to produce the bump (Iati et al. 2001).

The average observed interstellar extinction curve (Whittet, 2003)
is then compared with the model curves formed from a $\chi^2$
minimized and best fit linear combination of the composite and graphite
grains (for details see Vaidya \& Gupta 1999).

Figure 3(a) shows the interstellar extinction curve for the composite grains with AR=1.33
 (N=9640)
in the entire wavelength region of $3.4-0.10\mu m$ for the MRN grain size distribution, 
with the size range, a=0.005-0.250$\mu$. It is seen that the composite spheroidal
grain models with AR=1.33 (N=9640) and f=0.1 fits the average observed extinction
curve reasonably well in the wavelength range considered, i.e $3.4-0.10\mu m$.
The model extinction curves with AR=1.50 \& 2.00 (N=25896 \& 14440 respectively)
deviate from the observed
extinction curve in the uv region, beyond the wavelength
$\sim$ 0.1500 $\mu m$ (i.e. 6 $\mu m^{-1}$) and are thus not shown in the figure.
These results with the composite grains indicate
that the spheroidal grains with the axial ratio not very large
i.e AR $\sim 1.33$ (N=9640) is an optimum choice.
The results indicate that a third component of very small grains
(e.g very small silicate grains or PAHs) may be required to explain
the extinction beyond 1500\AA~ in the UV (Weingartner and Draine, 2001).

In the Figure 3(b), we have displayed the observed extinction curve in the direction
of the star HD46202 (data taken from IUE data base) and its best fitting with the model
AR=1.50 (N=25896) and grain size distribution of a=0.001-0.100$\mu$. We have selected
this particular star with $R_{v}=3.1$, from our recent analysis of extinction curves
towards the directions of 48 IUE stars (Katyal et, al., 2011)

\begin{figure}[t]
\centerline{\includegraphics[width=8.0cm,clip]{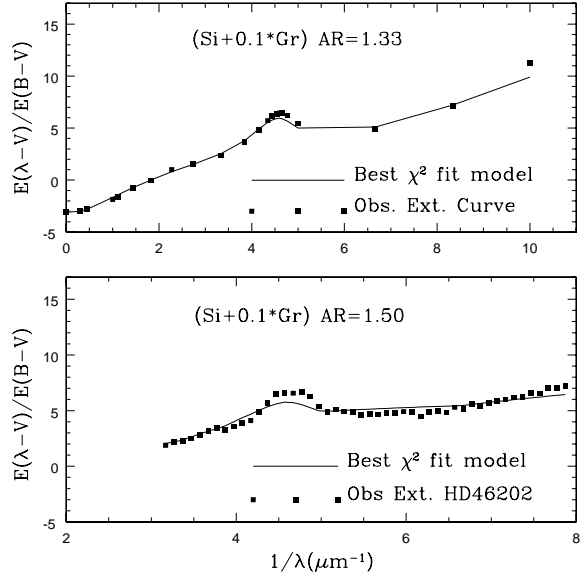}}
\caption{(a) Comparison of the observed interstellar extinction curve
(Whittet, 2003)
with the best fit model curve of composite grains with
graphite inclusions in the wavelength range of 3.4-0.1$\mu m$. (b) Observed Extinction
curve in the direction of the star HD46202 and its comparison with model curve of
AR=1.50.}
\end{figure}

Recently Iati et. al. (2004), Zubko et. al (2004)
Voshchinnikov et. al. (2005) and Maron \& Maron (2005)
have also proposed composite grain models. Very recently Voshchinnikov et. al (2006)
have proposed composite porous grain models with three or more grain populations and
have used both EMT-Mie type and layered sphere calculations.

\subsection{Cosmic Abundances}

In addition to reproducing the interstellar extinction curve, any
grain model must also be consistent with the abundance constraints.
Snow and Witt (1995, 1996) have reviewed several models for the
interstellar dust, which provide the data on the quantities of some
elements that are required to reproduce the interstellar extinction.
The number of atoms (in ppm) of the particular material tied up in grains
can be estimated if the atomic mass of the element in the grain material
and the density of the material are known
(see e.g. Cecchi-Pestellini et al. 1995 and Iati et al. 2001).
From the composite grain models we have proposed,
we estimate C abundance i.e. C/H between $\sim$ 165-200 (including those
atoms that produce the 2175\AA~ feature), which is
considerably lower than what is predicted by bare silicate/graphite grain models 
(e.g. C/H $\sim$ 254 ppm, Li and Draine, 2001) but it is still
significantly above the ISM value of $\sim$ 110 (Mathis 2000);
$\sim$ 140 (Sofia \& Meyer, 2001) and $\sim$ 100 (Sofia \& Parvathy, 2009).
The estimated Si abundance from the composite grain model presented
here is  between 25 and 30, which is lower than the other grain models,
32 ppm (Li and Draine, 2001) and is consistent with the recent ISM value of
25 ppm derived by Voshchinnikov \& Henning (2010). For appropriate reference
on abundance standards and related topics see Snow (2000) and Draine (2003).

\section{Summary and Conclusions}

Using the discrete dipole approximation (DDA)
we have studied the extinction properties of the composite
spheroidal grains, made up of the host silicate and graphite inclusions
in the wavelength region of 3.4-$0.10\mu m$.
Our main conclusions from this study are:

(1) The extinction curves for the composite spheroidal grains show a
shift in the central wavelength of the extinction peak as well as variation in
the width of the peak with the variation in the volume fraction of the
graphite inclusions. These results clearly indicate that the
shape, structure and inhomogeneity in the grains play an important role in 
producing the extinction. It must be noted here that large PAH molecules are also
candidates to the carrier of the interstellar 2175\AA~ feature --
a natural extension of the graphite hypothesis (Draine, 2003).

(2) The extinction curves for the composite spheroidal grains with the axial
ratio not very large (AR $\sim 1.33$, N=9640) and 10 \% volume
fractions of graphite inclusions are found to fit the average observed
interstellar extinction satisfactorily. 
Extinction curves with other composite grain models
with N=25896 and 14440 (i.e. with axial ratios of 1.50 and 2.00)
deviate from the observed curves in the UV region, i.e. beyond about wavelength 1500\AA~. 
These results indicate that a third component of very small particles in the composite 
grains may help improve the fit in the UV region (see e.g. Weingartner and Draine 2001).
It must be mentioned here that the composite spheroidal grain model with
silicate and graphite as constituent materials proposed by us is not
unique (see e.g. Zubko et al., 2004). We have also attempted to fit models to a specific
direction of the star HD46202 in our galaxy and show that AR=1.50 (N=25896) fits better
in this case. Analysis is in progress for many more such directions in the galaxy.

(3) These results clearly show that composite grain model is
more efficient, compared to bare silicate/graphite grain models,
in producing the extinction and it would perhaps help
reducing the cosmic abundance constraints.
Composite grain models with silicate, graphite and an additional
component (e.g. PAH's) may further reduce the abundance constraints.

We have used the composite spheroidal grain model to fit
the observed interstellar extinction and have derived the abundance of carbon (C/H)
and silicon (Si/H). The IRAS observations have indicated
the importance of the IR emission as a constraint on interstellar dust
models (Zubko et al. 2004). Recently, we have used the composite spheroidal
grain model to fit the IR emission curves obtained from IRAS observations
(Vaidya \& Gupta, 2011).

\acknowledgments{DBV thanks the organizing committee of the AOGS-2010,
for providing the opportunity to present the paper at meeting at
Hyderabad, India in July 2010. The authors acknowledge the financial support
from ISRO-Respond project (NO. ISRO/RES/2/2007-08). We would like to thank N V Voshchinnikov and the anonymous referee for their suggestions and useful comments on the manuscript.}

%%\lastpagecontrol{20cm}

\email
{\noindent
Nisha Katyal (nishakat@iucaa.ernet.in)\\
Ranjan Gupta (rag@iucaa.ernet.in)\\
D.B. Vaidya (e-mail: deepak.vaidya@iccsir.org)\\}
\label{finalpage}
\lastpagesettings
\end{document}